# Network structure of COVID-19 spread and the lacuna in India's testing strategy

Anand Sahasranaman [1,2], Nishanth Kumar [3]

[1]Division of Science, Division of Social Science, Krea University, Sri City, AP 517646, India.

[2]Centre for Complexity Science, Dept. of Mathematics, Imperial College London, London SW72AZ, UK.

[3]Independent Researcher.

**Abstract:**

We characterize the network of COVID-19 spread in India and find that the transmission rate is 0.43, with daily case growth driven by individuals who contracted the virus abroad. We explore the question of whether this represents exponentially decaying dynamics or is simply an artefact of India's testing strategy. Testing has largely been limited to individuals travelling from high-risk countries and their immediate contacts, meaning that the network reflects positive identifications from a biased testing sample. Given generally low levels of testing and an almost complete absence of testing for community spread, there is significant risk that we may be missing out on the actual nature of outbreak. India still has an apparently low current caseload, with possibly a small window of time to act, and should therefore aggressively and systematically expand random testing for community spread, including for asymptomatic cases. This will help understand true transmission characteristics and plan appropriately for the immediate future.

## 1. Introduction:

The novel coronavirus COVID-19 has rapidly spread across the world in the early months of 2020, with the World Health Organization categorizing it as a pandemic [1], leading to unprecedented measures to contain its spread [2]. India saw its first case of COVID-19 in a student who returned from Wuhan on January 30, 2020. After three initial cases in 2020 between January 30 and February 3, there were no further cases recorded for a month [3]. Since March 3, however, there has been a steady uptick of cases and there is a real concern of community spread occurring in India (Fig. 1a) [4]. There also appears to be heterogeneity in the spread of the virus across different states (Fig. 1b) [4].

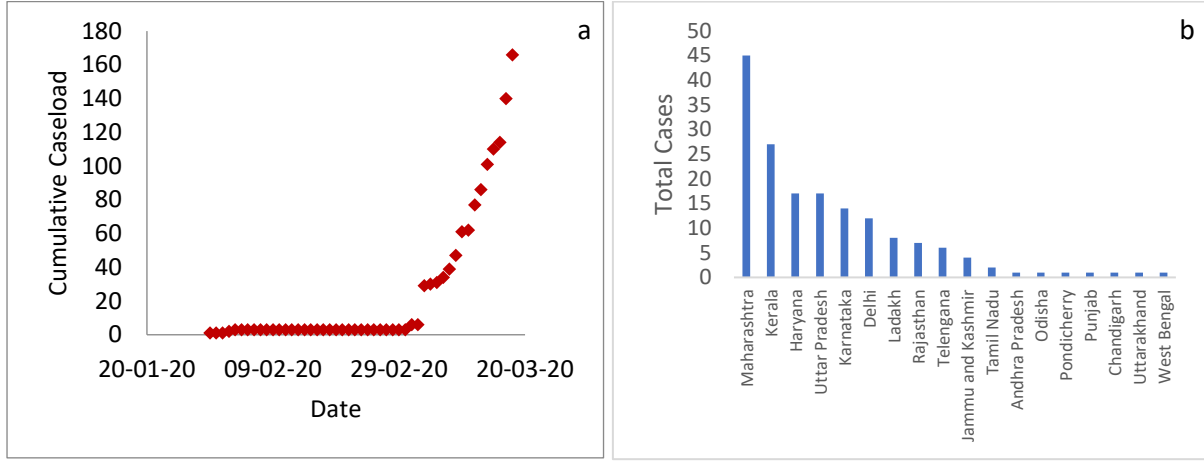

*Figure 1: The spread of COVID-19 in India (till March 18, 2020).* a) Change in cumulative caseload over time shows sharp rise of cases in March 2020. b) State-wise distribution of cases in India reveals Maharashtra to have the highest burden of cases.

The key parameter to characterize an outbreak is the transmission rate, which is a measure of the average number of individuals infected by an individual carrying the virus. A commonly used metric of transmission is the basic reproductive number $R_0$, which is the transmission rate given that the population has no immunity from past exposures or vaccination, nor any deliberate intervention in disease transmission [5]. When $R_0 > 1$, the number of infections grows and spreads in the population, but not for $R_0 < 1$. $R_0$ is generally modelled from data on the spread of the disease, with the assumption that the dynamic of spread is underway in the community and that the measure therefore parametrizes the extent of intra-community spread.

In this work, we look at the data on each individual case of COVID-19 infection and attempt to create the network structure of infections for India. Essentially, we attempt to identify how each individual was infected - whether the individual carried the infection from abroad (Level 1), or if it was local transmission in the country from an individual who had been infected abroad (Level 2), or if it was community transmission between individuals in the country (Level 3). Using this characterization of individual infections, we create the network of COVID-19 infections in India, and compute the transmission rate as the average number of neighbours of a node in the network. We also simulate different scenarios of transmission using a simple Susceptible-Infected-Recovered (SIR) epidemiology model [6], to highlight potential pathways of spread. The model considers a system of $N$ individuals split into 3 compartments at any given time $t$: $S(t)$ is the number of individuals susceptible to the infection; $I(t)$ is the number of infected individuals; and $R(t)$ is the number of recovered individuals. The model has two parameters: effective contact rate ($\beta$) and mean recovery rate ($\gamma$) and operates based on a set of three Ordinary Differential Equations: $\frac{dS}{dT} = -\frac{\beta SI}{N}$; $\frac{dI}{dT} = \frac{\beta SI}{N} - \gamma I$; $\frac{dR}{dT} = \gamma I$. Finally, we discuss implications for the strategy to contain COVID-19 in India.

## 2. The network of COVID-19 transmission:

We use March 13, 2020 as the date on which significant interventions happened in India to restrict the spread of disease, including the suspension of almost all visas for tourists entering the country, in addition to measures adopted by a number of states closing schools and colleges, as well as malls and movie halls. As of this date, we find that there were 86 cases of individuals who had contracted the virus, and we were able to identify the sources of infection in 82 of these cases. Because case numbers in India are still close to 200, we find that newspapers are still tracking individual case histories; and we use these newspaper reports and Wikipedia to construct the data set on individual transmission [3, 7-27]. Out of these cases, we identified 58 (71%) as Level 1, 23 (28%) as Level 2 or local transmission, and only one (1%) as Level 3 or community transmission (Fig. 2a).

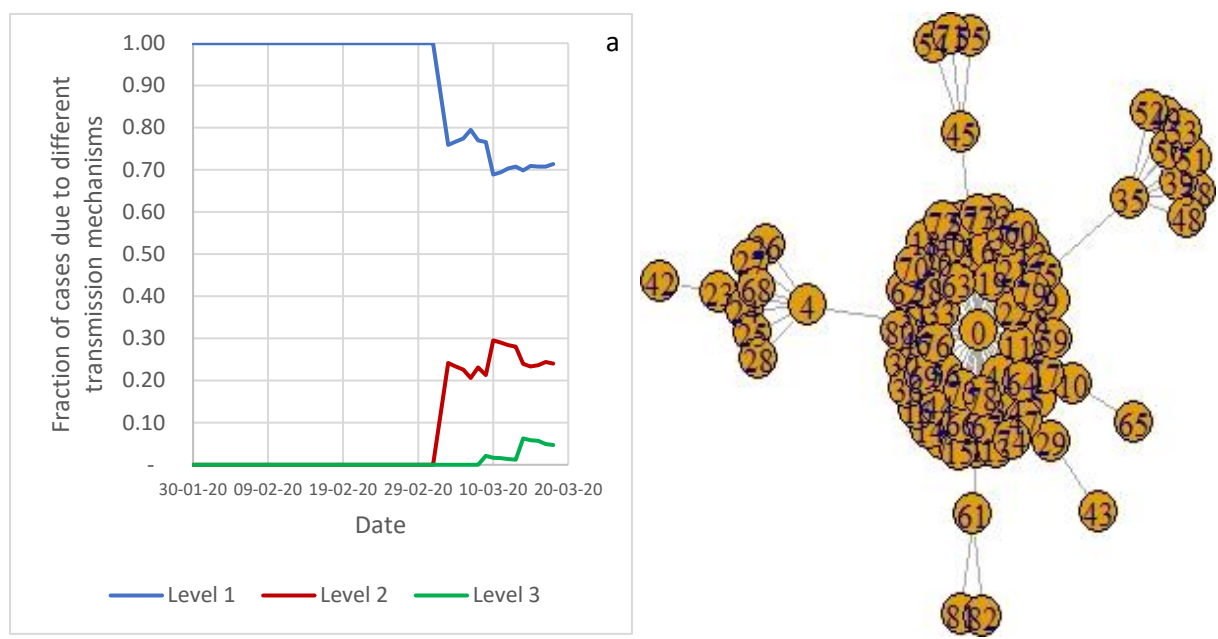

*Figure 2: The mechanisms of viral transmission.* a) Evolution of fraction of Level 1 (blue), Level 2 (red), and Level 3 (green) cases over time reveals that Level 1 remains the most significant transmission mode (until March 18, 2020). b) Network structure of COVID-19 infections in India until March 13, 2020 reveals that most transmission has occurred from individuals with recent travel history abroad (node 0).

Based on identification of the sources, we were able to create the network structure of the COVID-19 virus transmission (Fig. 2b), which shows that most cases of infection in India had their sources abroad (generalized as node 0), and a few of these infected nodes then contributed to local transmission. We computed the transmission rate of the virus, defined as the average number of neighbours of a node in the network (except node 0, which simply stands for infection prior to entering the Indian network), to be 0.43. This is equivalent to the transmission rate of the virus as it is a measure of the average number of infections caused by an infected individual. There appears to be an anomaly here with the revealed transmission rate of 0.43 ($< 1$) indicating that the spread of the virus should be decreasing, while the data clearly demonstrates that cases are going up from day-to-day (Fig. 1a). This anomaly is resolved when we assess the network structure, which reveals that a majority of infections still being identified are those who contracted the virus abroad (71%, Fig. 2), and this is a trend that continues in network structure even until March 18, when cumulative caseload was 166.

However, given the international travel restrictions put in place by the Government of India on March 13, we would expect that as their impact starts reflecting in the data by the end of March, and the number of new cases with Level 1 transmission will tend to zero. One might expect the overall spread of the disease to show an exponential decline from that point (as depicted in simulations in Figs. 3a and 3b), but before reaching any conclusions we must consider the all-important question of whether the data is a reasonable reflection of the true underlying dynamics of the spread of COVID-19 in India.

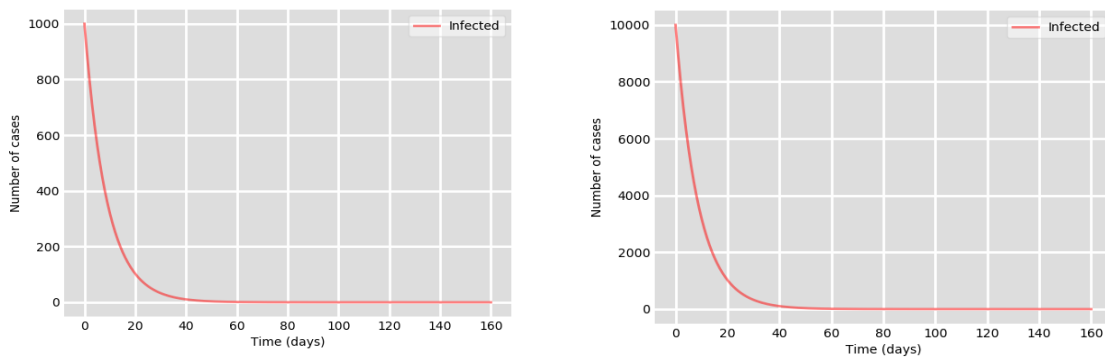

*Figure 3: SIR model of COVID-19 spread with transmission rate < 1.* a) Temporal evolution of caseload based on the following initial conditions and parameters: $N = S(0) = 1,300,000,000; I(0) = 1000; R(0) = 0; \beta = 0.086; \gamma = 0.2$. b) Temporal evolution of caseload based on the following initial conditions and parameters: $N = S(0) = 1,300,000,000; I(0) = 10,000; R(0) = 0; \beta = 0.086; \gamma = 0.2$.

**3. Issues in testing for community spread:**

In countries where the virus has seen community spread, $R_0$ has been estimated in the region of $2.5 - 3$, which indicates exponential growth dynamics [28]. While there is emerging research suggesting a degree of climate determination in the spread of COVID-19 as well as a correlation with geographical latitude, these results are very new and need confirmation [29,30]. Therefore, in the absence of clinching evidence relating to environmental factors, we ask the question of what is driving this discrepancy in transmission rates between India and other countries experiencing community spread.

To answer this question, we must first turn our attention to testing processes on the ground to identify cases of COVID-19 in India. It has been widely reported that the testing in India has been focused on (and limited to) citizens who have travelled from a set of high-risk countries and those who have come in contact with these cases [31-32]. There has been no systematic testing process in place for testing the extent of community spread in India [33]. Given this bias in testing, it should be no surprise that among the observed cases of COVID-19 in the country, a majority are travellers from high-risk countries and their immediate contacts with local transmission (as reflected in the networks structure of infections, Fig. 2b). Recently the Indian Council for Medical Research (ICMR) confirmed that limited testing on community spread had been initiated - 1,020 tests of people without travel history or contact with infected persons were to be done, though details of the randomization process in selecting these cases are not available [34]. While this is a start, it must be kept in mind that the actual network of spread will continue to reflect the same bias until community testing becomes a significant part of the testing strategy. Consolidated data from ICMR tells us that India has so far tested a total of 13,486 samples [35], or 10 tests people per million population, which is very low compared to other countries that have been testing for community spread [36], and creates the risk of missing such transmission in case it is already underway in the country. India currently has over 200 COVID-19 positive cases, but is yet to hit the kinds of caseload spikes seen in China, the US and Europe [2]. Given this shrinking window of opportunity and the significant lacuna in testing, India must, aggressively and systematically, proceed to widen the scope of testing for possible community transmission immediately. In this context, ICMR has just announced the introduction of tests for asymptomatic direct and high-risk contacts of a confirmed case [37]. However, given the tremendous void in testing for community spread so far, there is a need for larger scale random testing of both symptomatic and asymptomatic cases, in order to better understand the extent of spread in the broader community.

This is particularly critical because if indeed community transmission is underway in India, we could see significant proportions of the population being infected by COVID-19 before the epidemic recedes – we simulate these possibilities for $R_0 = 1.5$ (which is at the lower end of transmission rates from global evidence on COVID-19) and $R_0 = 2.5$ (which is close to the median transmission rate from evidence so far) and find that with community spread, we could see maximum infection caseloads of 6% and 23% of the population respectively (Figs. 4a and 4b).

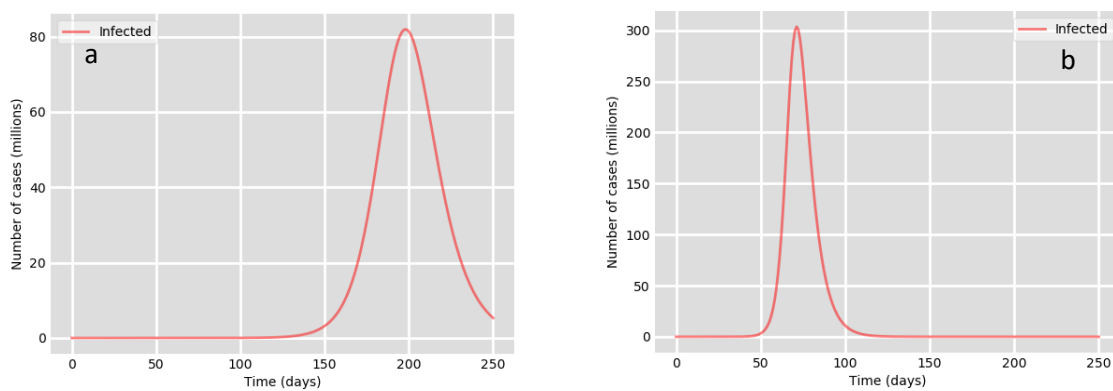

*Figure 4: SIR model of COVID-19 spread with transmission rate > 1.* a) Temporal evolution of caseload based on the following initial conditions and parameters: $N = S(0) = 1,300,000,000; I(0) = 1; R(0) = 0; \beta = 0.3; \gamma = 0.2$. b) Temporal evolution of caseload based on the following initial conditions and parameters: $N = S(0) = 1,300,000,000; I(0) = 1; R(0) = 0; \beta = 0.5; \gamma = 0.2$.

Epidemic spread models used by the UK and US predict much higher infection rates – with $R_0 = 2.4$, they estimate that up to 80% of the population of the US/UK would be infected in the absence of intervention [38]. Even if we assume that maximum infection load in India peaks at 6%, (about 80 million people) and even if only 5% of these cases are most severe requiring intensive care, this means that the Indian health system would need to cater for 4 million ICU beds, but it is estimated that there are only around 100,000 ICU beds in India – or 1 ICU bed for 40 people [39]. Given the state of the Indian health system, it is essential that we do as much as possible to test randomly for community spread – both symptomatic and asymptomatic - and attempt to unearth the gravity of spread in India as soon as possible, especially while the current restrictions on international travel, closure of places of gathering (schools, colleges, malls, temples, workplaces etc.), household quarantines, and other social distancing measures that are already in place.